\documentclass[preprint,amsmath,amssymb,aps,prl,preprintnumbers,nofootinbib]{revtex4-1}
\usepackage{graphicx}
\usepackage{color}

\def\ignore#1{{}}

\newcommand{\siml}{%
\hspace{0.3em}\raisebox{0.4ex}{$<$}\hspace{-0.75em}\raisebox{-.7ex}{$\sim$}\hspace{0.3em}}

\newcommand{\beeq}{\begin{equation}}
\newcommand{\eneq}{\end{equation}}
\newcommand{\beqn}{\begin{eqnarray}}
\newcommand{\eeqn}{\end{eqnarray}}

\def\dd{\partial}
\def\la{\raise.16ex\hbox{$\langle$}\lower.16ex\hbox{}  }
\def\ra{\raise.16ex\hbox{$\rangle$}\lower.16ex\hbox{} }
\def\go{\rightarrow}

\def\onehalf{ \hbox{$\frac{1}{2}$} }

\def\twothird{ \hbox{$\frac{2}{3}$} }

\def\eff{{\rm eff}}

\def\SM{{\rm SM}}
\def\EM{{\rm EM}}
\def\diag{{\rm diag ~}}

\def\KK{{\rm KK}}

\def\gammabar{ \bar{\gamma} }
\def\Gammabar{ \bar{\Gamma} }

\def\ep{\epsilon}
\def\psibar{ \psi \kern-.65em\raise.6em\hbox{$-$} }
\def\psibarl{ \psi \kern-.65em\raise.6em\hbox{$-$} \lower.6em\hbox{} }

\begin{document}

\title{Gauge-Higgs seesaw mechanism in six-dimensional grand unification}
\author{Yutaka Hosotani$^1$ and Naoki Yamatsu$^2$}
\affiliation{$^1$Department of Physics, Osaka University, Toyonaka, Osaka 560-0043, Japan\\
$^2$Maskawa Institute for Science and Culture, 
Kyoto Sangyo University, Kyoto 603-8555, Japan}
%
\date{7 August 2017}
\preprint{OU-HET 936,  MISC-2017-06}

\begin{abstract}
$SO(11)$ gauge-Higgs grand unification is formulated in the six-dimensional
hybrid warped space in which the fifth and sixth dimensions play as the electroweak
and grand-unification dimensions. 
Fermions are introduced in {\bf 32},  {\bf 11} and {\bf 1} of $SO(11)$.  
Small neutrino masses  naturally emerge as a result of
a new seesaw mechanism in the gauge-Higgs unification which is characterized by a $3 \times 3$ mass matrix.
\end{abstract}

\ignore{\qquad \small
}

\maketitle
\allowdisplaybreaks

The discovery of the Higgs boson at LHC supports the scenario of the 
unification of electromagnetic and weak forces.   In the standard model (SM)
the electroweak  (EW) gauge symmetry, $SU(2)_L \times U(1)_Y$, is spontaneously broken 
to $U(1)_\EM$ by the vacuum expectation value (VEV) of the Higgs scalar field.  
Although almost all observational data at low energies, including the data from 13 TeV LHC, 
are consistent with the SM, it is not clear whether the discovered Higgs boson is precisely 
what is introduced in the SM.  
The Higgs boson sector of the SM lacks a principle which governs and regulates the Higgs interactions 
with itself and other fields, in quite contrast to the gauge field sector which is controlled by the
gauge principle. 
At the quantum level the mass of the Higgs boson $m_H$ acquires large corrections
which have to be canceled by fine-tuning a bare mass in the theory.
It is called the gauge-hierarchy problem.

There are many proposals  to overcome these problems.
One possible  scenario is  the gauge-Higgs unification in which the Higgs
boson is identified with a part of the extra-dimensional component of gauge fields defined
in higher dimensional spacetime.\cite{YH1, Davies1, Hatanaka1998}
The Higgs boson appears as a four-dimensional fluctuation mode of the Aharonov-Bohm (AB) 
phase $\theta_H$ along the extra-dimensional space.
It acquires  a finite mass at the quantum level, independent of the cutoff scale. 

In the EW interactions the $SO(5) \times U(1)$ gauge-Higgs  unification 
in the five-dimensional  Randall-Sundrum (RS)  warped space has been 
formulated.\cite{ACP2005, Medina2007, HS2007,FHHOS2013} 
It has been shown to give  almost the same phenomenology at low energies as the SM
for $\theta_H \siml 0.1$.
It gives many predictions to be explored and confirmed in the coming experiments at LHC 
and in future experiments at $e^+ e^-$ colliders.
For instance, it predicts the $Z'$ bosons (the first KK modes of $\gamma$, 
$Z$ and $Z_R$) and  $W'$ boson  (the first KK modes of  $W$)  
around 7 to 8 TeV range for $\theta_H = 0.1$ to $0.07$, and
larger forward-backward asymmetry in $e^+ e^-$ collisions at 250$\,$GeV$\sim 1\,$TeV
than in the SM.

As a next step  it is natural  to incorporate strong interaction to achieve
gauge-Higgs grand unification (GHGU).  
The mere fact of  charge quantization in the quark-lepton spectrum
strongly indicates  grand unification.  Such attempts have been already 
made.\cite{Burdman2003, Lim2007, Yamashita2011, Serra2011, HY2015a, Yamatsu2016a, Furui2016}
Recently the $SO(11)$ gauge-Higgs grand unification model in the five-dimensional
Randall-Sundrum warped space has been proposed, in which the EW Higgs boson emerges 
from the fifth dimensional component of the gauge potentials and many good features of
the $SO(5) \times U(1)$ gauge-Higgs EW unification are carried over.\cite{HY2015a, Furui2016}
The breaking of the symmetry of grand unification is achieved there by imposing different orbifold 
boundary conditions at the UV and IR branes in the RS space.  On the UV brane $SO(11)$ is
broken to $SO(10)$, whereas on the IR brane  to $SO(4) \times SO(7)$.
As a result the remaining gauge symmetry becomes $SO(4) \times SO(6)$.
The brane scalar in {\bf 16} of $SO(10)$ is introduced on the UV brane, which
spontaneously breaks $SO(10)$ to $SU(5)$, leaving the SM gauge symmetry as a whole.
It is found that  proton decay is strictly forbidden in the minimal model.  
The mass spectrum of quarks and leptons is realized in the combination of
the Hosotani mechanism and $SO(10)$-invariant interactions  on the UV brane.

However, as a consequence of the two distinct orbifold boundary conditions imposed 
on the UV and IR branes, there necessarily emerge light exotic fermions ($\hat u$) 
of charge $- \twothird e$.  $\hat u$ has parity either $(+,-)$ or $(-, +)$.  
It does not cause a problem in flat space, but in the RS space its mass $m_{\hat u}$
turns out to be about $m_u \cot \onehalf \theta_H$, which contradicts with the
observation for $\theta_H \sim 0.1$.
The problem is unavoidable in the five-dimensional RS space.\cite{Furui2016}
Furthermore it is difficult to naturally explain  small masses of neutrinos.

To overcome these difficulties we  propose gauge-Higgs
grand unification in the six-dimensional hybrid warped space.  
Consider six-dimensional spacetime with a metric
\begin{align}
ds^2 &= e^{- 2 \sigma (y)} ( \eta_{\mu\nu} dx^\mu dx^\nu + dv^2 ) + dy^2 
\label{metric1}
\end{align}
where $\eta_{\mu\nu} = \diag (-1, 1,1,1)$,  
$\sigma(y) = \sigma(y + 2 L_5) = \sigma (-y)$, and $\sigma(y) =k |y|$ for$|y| \le L_5$.
We identify spacetime points
$(x^\mu, y, v)$, $(x^\mu, y+ 2 L_5, v)$,  $(x^\mu, y, v + 2\pi R_6)$, and $(x^\mu, -y, -v)$.
The spacetime has the same topology as $M^4 \times (T^2/Z_2)$.
It naturally allows to have chiral fermions in four dimensions.
There appear four fixed points in the extra-dimensional space under $Z_2$
parity; $(y_0, v_0) = (0,0)$, $(y_1, v_1) = (L_5, 0)$, $(y_2,v_2) = (0, \pi R_6)$
and $(y_3, v_3) = (L_5, \pi R_6)$.  Parity $P_j$ around each fixed point is
defined by $ (x^\mu, y_j + y , v_j + v)  \go  (x^\mu, y_j - y , v_j - v) $.
The metric \eqref{metric1} solves the Einstein equations with
the brane tension at $y=0$ and $L_5$ and a negative cosmological constant
$\Lambda = - 10 k^2$.  There are five-dimensional branes at $y=0$ and $L_5$, 
each of which has topology of $M^4 \times S^1$.
The spacetime \eqref{metric1} generalizes the RS space, and is called as
the hybrid warped space hereafter.

There appear two Kaluza-Klein mass scales
$m_{ \KK_5} =  \pi k/( e^{kL_5} -1)$ and $m_{ \KK_6} = R_6^{-1}$
in the fifth and sixth dimensions.
We suppose that the warp factor is large; $z_L = e^{kL_5} \gg 1$.  
$m_{ \KK_5}$ turns out to be $6 \sim 10 \,$TeV as in the $SO(5) \times U(1)$
gauge-Higgs EW unification.  $m_{ \KK_6}$ is expected to be a GUT scale,
and therefore $m_{ \KK_6} \gg m_{ \KK_5}$.
Not all $P_j$'s are independent.  It is easy to see that
$P_3 = P_2 P_0 P_1 = P_1 P_0 P_2$.  Further loop translations along
the fifth and sixth dimensions, 
$U_5$: $(x^\mu, y, v) \go (x^\mu, y + 2 L_5, v)$ and 
$U_6$: $(x^\mu, y, v) \go (x^\mu, y , v + 2\pi R_6)$, are related by
$U_5 = P_1 P_0 = P_3 P_2$  and $U_6 = P_2 P_0 = P_3P_1$.

We consider $SO(11)$ gauge theory in the hybrid warped space.
Gauge potentials satisfy 
\beeq
\begin{pmatrix} A_\mu \cr A_y \cr A_v \end{pmatrix} (x, y_j -y, v_j -v)
= P_j \begin{pmatrix} A_\mu \cr - A_y \cr - A_v \end{pmatrix} (x, y_j +y, v_j +v) P_j^{-1}
\label{gaugeBC1}
\eneq
where $P_j {\rm ~or~} - P_j \in SO(11)$ and $(P_j)^2 = 1$.  
We take, in the vectorial representation, 
$P_0^{vec} =P_1^{vec} = I_4 \oplus (- I_7)$ and 
$P_2^{vec} = P_3^{vec} = I_{10} \oplus (-I_1)$.
Note $U_5 = P_1 P_0 = 1$.
The choice $P_0=P_1$ and $P_2=P_3$ enables us to avoid light exotic particles
in the warped space.  $SO(11)$ is reduced to $SO(4) \times SO(7)$ by $(P_0, P_1)$, 
and to $SO(10)$ by $(P_2, P_3)$.
As in the 5d model, the orbifold boundary conditions reduce $SO(11)$ to
$SO(4) \times SO(6)$.  In the representation of the $SO(11)$ Clifford algebra 
in Ref.\ \cite{Furui2016}, the corresponding $P_j$'s  in the spinorial representation are
given by
$P_0^{sp} = P_1^{sp} = I_2 \otimes \sigma^3 \otimes I_8$ and
$P_2^{sp} = P_3^{sp} =  I_{16} \otimes \sigma^3$.

Four fermion multiplets in the spinor representation, $\Psi_{\bf 32}^\alpha (x,y,v)$
($\alpha = 1 \sim 4$), are introduced.  Three of them ($\alpha=1,2,3$) contain three 
generations of quarks and leptons.  
Dirac matrices in 6 dimensions satisfy $\{ \Gamma^a, \Gamma^b \} = 2 \eta^{ab}$
($\eta^{ab} = \diag (-1,1, 1, 1, 1, 1)$).
6d chirality matrix is given by $\gamma^7_{6d} = I_4 \otimes \sigma^3$,  which is 
related to 4d chirality matrix $\gamma^5_{4d} = I_2 \otimes \sigma^3 \otimes I_2$ by 
$\gamma^5_{4d} \gamma^7_{6d}  = \gamma^7_{6d} \gamma^5_{4d}
= -i \Gamma^5 \Gamma^6 \equiv \gammabar$.
The orbifold boundary conditions are
\beeq
\Psi_{\bf 32}^\alpha (x,y_j - y , v_j - v) = \eta_j^\alpha \gammabar P_j^{sp}
\Psi_{\bf 32}^\alpha (x,y_j + y , v_j + v)
\label{fermionBC1}
\eneq
where $\eta_j^\alpha = \pm 1$.  We impose the 6d Weyl condition such that
$\gamma^7_{6d} = +1 \, (-1)$ for $\alpha=1,2 \, (3,4)$, to ensure the cancellation
of 6d chiral anomaly.  
In the current representation the upper (lower) half of  $\Psi_{\bf 32}$ corresponds to 
${\bf 16}$ ($\overline{\bf 16}$) of $SO(10)$.
We choose $\eta_j^{1,2} = -1$, $\eta_j^{3} = 1$, and 
$\eta_{0,2}^{4} = - \eta_{1,3}^{4} = 1$.
One finds that
the three generations ($\alpha=1,2,3$) have zero modes corresponding to quarks and leptons
such that all left-handed $SU(2)_L$ doublets are in ${\bf 16}$ and all right-handed $SU(2)_L$ 
singlets are in $\overline{\bf 16}$.  The zero mode structure is the same as in the 5d model of
Refs. \cite{HY2015a, Furui2016}.  The exotic particle components encountered in the 5d model,
denoted by the $\hat{~}$ symbol there, have all $P_0 =P_1 = - P_2 =- P_3$.
Fields with $P_0 =P_1 = - P_2 =- P_3$ are expanded in Fourier series of either $\cos (n+\onehalf)v/R_6$
or  $\sin (n+\onehalf)v/R_6$ so that the mass of each mode is equal to or greater than 
$1/2 R_6 = \onehalf m_{\KK_6} \gg m_{\KK_5}$.   
The problem of the appearance of light exotic fermions in the 5d GHGU is solved.
The fourth generation $\Psi_{\bf 32}^{4}$ does not have any zero mode.
Its lightest mode has a mass of order of $m_t \cot \onehalf \theta_H$.
$\Psi_{\bf 32}^{4}$ plays the role of dark fermions in the $SO(5) \times U(1)$ gauge-Higgs 
EW model, and is necessary to have 6d anomaly cancellation as well.

Dirac fermions  in the vector representation, $\Psi_{\bf 11}^{\beta}$ and $\Psi_{\bf 11}^{\prime \beta}$
($\beta=1,2,3$),   are also introduced in the 6d bulk.
The boundary conditions are given by formulas similar to (\ref{fermionBC1}) with 
$\eta_j^\alpha \gammabar P_j^{sp}$ replaced by $\eta \gammabar P_j^{vec}$ where
$\eta = + (-)$  for $\Psi_{\bf 11}^{\beta}$ ($\Psi_{\bf 11}^{\prime \beta}$).
The fermion spectrum at low energies is the quark-lepton spectrum in the SM.

There are brane fields defined on the UV brane at $y=0$.
A single brane scalar field in the spinor representaion of $SO(11)$, $\Phi_{\bf 32}  (x, v)$, 
is introduced.  Its VEV,  $ w/\sqrt{\pi R_6}$,  spontaneously breaks 
$SO(11)$ to $SU(5)$.   As a result $SO(11)$ is reduced to the SM symmetry, 
${\cal G}_\SM= SU(2)_L \times SU(3)_C \times U(1)_Y$.  
One needs to assume only $w \gg m_{\KK_5}$.
We note that  the $SO(11)$ gauge invariance on the UV brane is demanded as the brane covers 
the bulk region $0< v < \pi R_6$.
$\Phi_{\bf 32} $ satisfies the boundary condition
$\Phi_{\bf 32} (x, v_j - v) = \eta_j  P_0^{sp} \Phi_{\bf 32} (x, v_j + v)$ ($j=0,2$) 
where $(\eta_0, \eta_2) = (-,+)$.
The $SU(5)$ singlet component of {\bf 16} in $SO(10)$ has a zero mode and 
develops nonvanishing VEV.

In addition to the SM gauge fields,  $A_y^{a \, 11}$ and $A_v^{a \, 11}$ ($a=1 \sim 4$) 
have zero modes.  The zero modes of $A_y^{a \, 11}$ ($a=1 \sim 3$) are absorbed by 
the $W$ and $Z$ bosons.  The zero mode of $A_y^{4 \, 11}$ becomes the 4d Higgs
field.  Its mass, $m_H =125\,$GeV,  is generated at one loop.
The zero modes of $A_v^{a \, 11}$ ($a=1 \sim 4$) are 4d scalars with masses of 
$O(g_4 m_{\KK_6})$ generated at the one loop level.
Eight components of complex $\Phi_{\bf 32}$ have zero modes under  the boundary conditions. 
Among them the $SU(5)$ singlet component acquires nonvanishing VEV.  
Nine components out of the 16 real components are absorbed by
$SO(4)\times SO(6)/{\cal G}_\SM$ gauge fields.  
The remaining 6 ($= 16 -1-9$) pseudo-Nambu-Goldstone bosons acquire masses of $O(g_4 w)$ at the
one loop level.

The action of a fermion field  $\Psi$ in 6d bulk is given by
$\int d^6 x  \sqrt{- \det G}  \, \overline{\Psi} \big\{ \Gamma^a {E_a}^M \big( D_M  
+ \frac{1}{8} \Omega_{bcM} [\Gamma^b , \Gamma^c ] \big) + i c \sigma'(y) \Gamma^6 \big\} \Psi$ 
where 
$ \overline{\Psi} = i \Psi^\dagger \Gamma^0$, $D_M = \dd_M - ig A_M$, 
and $\Omega_{bcM}$'s are spin-connections.
The last term with the coefficient $c$ 
represents a bulk vector mass in the hybrid warped space,
which generalizes a bulk scalar mass  in the RS space.
For $\Psi_{\bf 11}$  an additional  mass term 
$m_{\bf 11} \overline{\Psi}_{\bf 11} \Psi_{\bf 11}$ is allowed.
It is most convenient to work  in the conformal coordinate $z= e^{ky}$ in the 
fundamental region $1 \le z \le z_L , 0 \le v < 2 \pi R_6$.
For a 6d Weyl fermion with $\gamma_{6d}^7 = +$, we write
$\Psi = z^{5/2} (\xi, \eta, 0 ,0)$ where $\xi, \eta$ are two-component 
right- and left-handed spinors.
The action becomes
\begin{align}
& \int d^4 x \int_0^{2\pi R_6} dv \int_1^{z_L} \frac{dz}{k} \,  i (- \eta^\dagger, \xi^\dagger )  \cr
&
\times \begin{pmatrix}
- k \hat D_-(c) + i D_v & \sigma^\mu D_\mu \cr
\bar \sigma^\mu D_\mu & - k \hat D_+ (c) + i D_v 
\end{pmatrix}
\begin{pmatrix} \xi \cr \eta \end{pmatrix} 
\label{fermionAction2}
\end{align}
where $\hat D_\pm (c) = \pm D_z + (c/z)$.
We stress that the bulk vector mass term $\overline{\Psi} i c \sigma' (y) \Gamma^6 \Psi$
precisely plays the role of the bulk scalar mass in the 5d GHGU in the RS space.

The parity $P_0=P_2=+1$ components of the 
6d bulk fermion fields, $\Psi_{\bf 32}^\alpha$, $\Psi_{\bf 11}^\beta$ and 
$\Psi_{\bf 11}^{\prime \beta}$, have brane interactions with the brane scalar 
$\Phi_{\bf 32}$ on the UV brane at $y=0$.  
They take the $SO(11)$ invariant form such as
$\delta(y) \overline{\Psi}_{\bf 32} \Gamma^a \Phi_{\bf 32} (\Psi_{\bf 11})_a$.  
With $\la \Phi_{\bf 32} \ra \not= 0$ they generate mass mixing among $\Psi_{\bf 32}$
and $\Psi_{\bf 11}$.  It can be shown that the observed mass spectrum of quarks and charged leptons 
is reproduced in the combination of the Hosotani mechanism, brane interactions and 
$\overline{\Psi}_{\bf 11} \Psi_{\bf 11}$ terms,  which will be reported separately.  
In the present paper we focus on the neutral fermion sector, 
and show how small neutrino masses are generated by a new seesaw mechanism.

$SO(11)$ singlet, eight-component brane fermions $\chi^\beta (x,v)$ ($\beta=1 \sim 3$)
are introduced on the UV brane for three generations.  $\chi^\beta$ satisfies
$\chi^\beta (x, v_j - v) = \gammabar \,  \chi^\beta  (x, v_j + v)$ ($j=0,2$).
For the sake of simplicity in notation we suppress the generation index $\alpha, \beta$ hereafter.
With $\chi$ written in terms of two-component spinors as $\chi = (\xi_+, \eta_+, \xi_-, \eta_-)$,
the orbifold boundary condition  implies that only $\xi_+$ and $\eta_-$ have
zero modes ($v$-independent modes).
Recall that charge conjugation $C$ in six dimensions is given by
$\chi^C = e^{i\delta_C} (- \sigma^2 \eta_+^*, - \sigma^2 \xi_+^*, \sigma^2 \eta_-^*, \sigma^2 \xi_-^*)$.
$\chi$ lives on the five-dimensional brane so that one can impose the simplectic Majorana 
condition \cite{Peskin1998} on $\chi$, 
\begin{align}
\chi^C =  \tilde \chi \equiv  i \Gammabar \chi ~, 
\label{chargeconjugate2}
\end{align}
where $\Gammabar = \gamma_{4d}^5 \Gamma^6$.
As $\Gammabar$ commutes with $\Gamma^a$ ($a=0,1,2,3,6$) and  anti-commutes with $\Gamma^5$,
$\tilde \chi$ transforms as a 5d spinor.
One finds that $\xi_+ = \eta_-^c = - e^{i \delta_C} \sigma^2 \eta_-^*$ and 
$\eta_+ = \xi_-^c = e^{i \delta_C} \sigma^2 \xi_-^*$.  
Combined with the boundary conditions, 
$\chi$ has only one independent  zero mode in $\eta_-$.

For $\chi$ field a mass term $\onehalf M \overline{\chi} \chi \delta (y)$ is allowed.
Further there is a brane interaction 
$- \delta(y) \kappa_B \{ \overline{\chi} \Phi_{32}^\dagger \Psi_{\bf 32} + h.c. \}$, which generates,
with $\la \Phi_{32} \ra \not= 0$, mass mixing  among $\chi$ and $\nu'$ where $(\nu' , e')$ is 
an $SU(2)_L$-singlet and $SU(2)_R$-doublet in $\Psi_{\bf 32}$.\cite{Furui2016}
As a result the action for $\chi$ is given by $\int d^4x  \int_0^{2\pi R_6} dv ({\cal L}_k + {\cal L}_m)$
where ${\cal L}_k = \onehalf \overline{\chi} (\Gamma^\mu \dd_\mu + \Gamma^6 \dd_v) \chi$ and 
\begin{align}
{\cal L}_m =
- \frac{m_B}{\sqrt{k}} (\overline{\chi} \nu' + \overline{\nu}' \chi ) |_{y=0} 
- \onehalf M \overline{\chi} \chi  ~.
\label{chiAction1}
\end{align}
The relevant fields determining the observed neutrino are 
$\chi$, $\nu'$ and $\nu$ of $(\nu, e)$ which is an $SU(2)_L$-doublet and $SU(2)_R$-singlet  
in $\Psi_{\bf 32}$.
Zero modes are contained in $\nu_L$, $\nu_R'$ and $\eta_-$.
We would like to stress that $M \overline{\chi} \chi $ gives rise to 
$M( - i \eta_-^\dagger \eta_-^c + i \eta_-^{c\dagger} \eta_-)$, 
namely a Majorana mass in four dimensions.

Due care must be taken in deriving mass eigenvalues, as $\nu$ and $\nu'$ are 6d fields whereas
$\chi$ is a 5d field.  Before giving  rigorous treatment, it is instructive to present an effective 
theory in four dimensions.  
Let $(\nu_{0L}, \nu_{0R}', \eta_{0L})$ be canonically normalized 
4d fields associated with the zero modes of $(\nu, \nu', \chi)$.
In perturbative expansion  $\eta_- (x,v) = (2\pi R_6)^{-1/2} \eta_{0L} (x) + \cdots$ and 
$\nu_R' (x, z, v)= (k/2\pi R_6)^{1/2} z ^{5/2} u_R^{(0)} (z; c_{\bf 32}) \nu_{0R}'(x) + \cdots$ 
where $c_{\bf 32}$ is the bulk vector mass parameter of $\Psi_{\bf 32}$ and 
the mode function $u_R^{(0)} (z; c)$ is normalized by
$\int_1^{z_L} dz u_R^{(0)} (z; c)^2 =1$. 
With $\theta_H \not= 0$, $\nu$ and $\nu'$ mix by the Hosotani mechanism, generating
an effective Dirac mass 
$m_D (i \nu_{0L}^\dagger \nu_{0R}' - i \nu_{0R}^{\prime \dagger} \nu_{0L})$ 
just as other quark and lepton pairs do.
Then the mass terms are written as
\begin{align}
\frac{i}{2} (\nu_{0L}^{c  \, \dagger},  \nu_{0R}^{\prime \, \dagger}, \eta_{0L}^{c \, \dagger} )
\begin{pmatrix} & m_D & \cr m_D && \tilde m_B \cr &\tilde m_B &M \end{pmatrix} 
\begin{pmatrix} \nu_{0L} \cr \nu_{0R}^{\prime c} \cr \eta_{0L} \end{pmatrix} + h.c. 
\label{mass1}
\end{align}
where $\tilde m_B = m_B u_R^{(0)} (1; c_{\bf 32})$.
$u_R^{(0)} (z; c)$ is given by $(2c+1)^{1/2} z_L^{- 1/2} (z/z_L)^c$ in free theory.
Mass eigenvalues $\kappa$ are determined by 
$\kappa^3 - M \kappa^2 - (\tilde m_B^2 + m_D^2) \kappa + m_D^2 M =0$.
In particular, for $m_D \ll M ,  \tilde m_B$, one finds
$\kappa_{\rm large} \sim \onehalf \big\{ M \pm \sqrt{M^2 + 4 \tilde m_B^2} \big\} $ and
\begin{align}
\kappa_{\rm small} \sim \frac{m_D^2 M}{\tilde m_B^2} ~,
\label{mass2}
\end{align}
which is identified as the small neutrino mass.
This should be compared with the seesaw mechanism in four dimensions \cite{Minkowski1977}
which is typically characterized by a $2 \times 2$ matrix in each generation.
Note that the Majorana mass $M$ appears in the numerator of  (\ref{mass2}).
In the gauge-Higgs grand unification, an effective Majorana mass $\tilde m_B^2/M$ for $\nu_R'$
is induced through $\nu_R' \go \eta_L \go \eta_L^c \go \nu_R^{\prime c}$.

To confirm the above picture and derive a precise formula, 
we return to the equations in the six-dimensional hybrid warped space.
Each field is expanded in a Fourier series in the sixth coordinate $v$.  Only fields with 
$P_0 P_2 = P_1 P_3 = +1$ have zero modes.  
All other modes have large masses $\ge \onehalf m_{\KK_6}$.
Supposing that $\tilde m_B, M \ll m_{\KK_6}$, we safely retain only $v$-independent modes.  
Further we assume, for the sake of simplicity, that the mixing of $\nu$ and $\nu'$ with 
other neutral component of $\Psi_{\bf 32}$ and $\Psi_{\bf 11}$,  induced by
brane interactions, may be neglected.  
After factoring out $z^{5/2}$ for $\nu$ and $\nu'$ as done in (\ref{fermionAction2}), one finds
\begin{align}
&-k \hat D_-  (c_{\bf 32}) \begin{pmatrix} \nu_R \cr \nu_R' \end{pmatrix} 
+ \sigma  \dd \begin{pmatrix} \nu_L \cr \nu_L' \end{pmatrix} = 0 ~, \cr
&\bar \sigma \dd \begin{pmatrix} \nu_R \cr \nu_R' \end{pmatrix}
- k \hat D_+  (c_{\bf 32}) \begin{pmatrix} \nu_L \cr \nu_L' \end{pmatrix} 
= \delta (y) \begin{pmatrix} 0 \cr m_B \eta_L / \sqrt{k} \end{pmatrix} , \cr
&\Big\{ \sigma \dd \, \eta_L - \frac{m_B}{\sqrt{k}}  \nu_R' - M \eta_L^c \Big\} \delta (y) = 0 ~.
\label{equation1}
\end{align}
Here $- g A_z = \frac{1}{2} \theta' (z) \tau_1$ in $\hat D_\pm (c_{\bf 32})$
where $\theta (z) =\theta_H (z_L^2 - z^2)/(z_L^2 -1)$.  
$\nu_L$ and $\nu_R'$ are parity even at $y=0$ and $L_5$, whereas $\nu_R$ and $\nu_L'$ are parity odd.
It follows from (\ref{equation1}) that at $y=\ep$ ($z=1^+$)
\begin{align}
&\nu_R (x, 1^+) = 0 ~, \cr
&\hat D_+ \nu_L (x, 1^+)  = D_+ \nu_L (x,1) + \frac{i}{2} \theta' (1^+) \nu_L'  (x,1^+) = 0 ~, \cr
&- 2 \nu_L' (x, 1^+)  = \frac{m_B}{\sqrt{k}} \eta_L (x) ~, \cr
& k  D_- \nu_R' (x, 1^+)  
= - \frac{m_B}{2\sqrt{k}} \Big\{ \frac{m_B}{\sqrt{k}} \nu_R' (x, 1) + M \eta_L^c (x) \Big\} 
\label{BC1}
\end{align}
where $D_\pm = \pm (\dd/\dd_z) + (c_{\bf 32}/z)$.
$\nu_L'$ develops a discontinuity at $y=0$ due to the brane interaction with $\eta_L$.

At this stage it is convenient to  move to the twisted gauge in which 
$(\tilde \nu,\tilde \nu')^t = \exp \big\{ \frac{i}{2} \theta (z) \tau_1 \big\}  (\nu , \nu' )^t$ and $\tilde A_z = 0$.  
$\tilde \nu$ and $\tilde \nu'$ satisfy free equations in the bulk $1 < z \le z_L$.
The boundary conditions at $z=z_L$ remain unchanged, and are given by
$D_+ \tilde \nu_L = \tilde \nu_R = \tilde \nu_L' = D_- \tilde \nu_R' =0$.
As a consequence solutions to (\ref{equation1}) satisfying the boundary conditions at $z_L$ can be 
expressed, in terms of basis functions  defined in  Appendix B of Ref.\ \cite{Furui2016},  as
\begin{align}
\begin{pmatrix} \tilde \nu_R \cr \tilde \nu_R' \end{pmatrix} &= \sqrt{k} 
\begin{pmatrix} \alpha_\nu S_R (z; \lambda, c) \cr 
     i \alpha_{\nu '} C_R (z; \lambda, c) \end{pmatrix} f_R(x) ~, \cr
\begin{pmatrix} \tilde \nu_L \cr \tilde \nu_L' \end{pmatrix} &= \sqrt{k}
\begin{pmatrix} \alpha_\nu C_L (z; \lambda, c) \cr 
    i \alpha_{\nu '} S_L (z; \lambda, c) \end{pmatrix} f_L(x) ~, \cr
\eta_L &= i  \alpha_\eta f_L (x) ~,
\label{mode1}
\end{align}
where $c = c_{\bf 32}$,  
$D_+ (C_L, S_L)=\lambda (S_R, C_R)$, $D_-(C_R, S_R) = \lambda (S_L, C_L)$,  and
$C_L C_R - S_L S_R=1$.
$C_L=C_R=1$ and $S_L=S_R=0$ at $z=z_L$.
$f_{R,L}$ satisfy 
$\sigma \dd f_L = k \lambda f_R$, $\bar \sigma \dd f_R = k \lambda f_L$ and $f_L ^c = f_R $.
$\alpha_\eta$ is taken to be real.

It follows from (\ref{BC1}) that 
$K (\alpha_\nu ,  \alpha_{\nu '} ,  \alpha_\eta)^t =0$ where 
\begin{align}
K = \begin{pmatrix} \cos \onehalf\theta_H S_R & \sin  \onehalf\theta_H C_R & 0 \cr
-  \sin  \onehalf\theta_H C_L &  \cos \onehalf\theta_H S_L &  \onehalf m_B/k  \cr
-  \sin  \onehalf\theta_H S_R & \cos \onehalf\theta_H C_R & -(k\lambda + M)/m_B 
\end{pmatrix} .
\label{BC2}
\end{align}
Here  $S_R = S_R(1; \lambda, c)$ etc.  
Mass eigenvalues $k\lambda$ are determined by $\det K =0$;
\begin{align}
(k \lambda + M) \Big\{ S_L S_R + \sin^2 \onehalf \theta_H \Big\} 
+ \frac{m_B^2}{2k} S_R C_R = 0 ~.
\label{mass3}
\end{align}
For  $\lambda z_L \ll1$ and $k \lambda \ll |M|$ one finds that
\begin{align}
\lambda = - \frac{2k M}{m_B^2} \sin^2 \onehalf \theta_H   
\begin{cases} (1- 2c) z_L^{2 c -1} &{\rm for} ~ c < \onehalf , \cr
2 c -1 &{\rm for} ~ c > \onehalf .\end{cases}
\label{mass4}
\end{align}
The Dirac mass $m_D$ acquired through the Hosotani mechanism in the neutral sector 
is the same  as the  mass  $k \lambda_D$ of the up-type quark
determined by $S_L S_R + \sin^2 \onehalf \theta_H =0$.\cite{Furui2016}
One finds that $\lambda_D z_L = \sqrt{1 - 4c^2}  \sin \onehalf \theta_H$ for $c<\onehalf$ 
and  $ \sqrt{4c^2-1} z_L^{-c+(1/2)} \sin \onehalf \theta_H$ for $c>\onehalf$.
We recall that $c > \onehalf$ ($< \onehalf$) for $u, c$ ($t$) quarks. %
Thus the small neutrino mass is given by the gauge-Higgs seesaw formula
\begin{align}
m_\nu = - \frac{m_D^2 M z_L^{2c +1}}{(c +\onehalf)  m_B^2 } ~.
\label{mass5}
\end{align}
If the  estimate $u_R^{(0)} (1; c) \sim (2c+1)^{1/2} z_L^{- 1/2-c} $ in free theory
were inserted into $\tilde m_B$ in the formula (\ref{mass2}) in the 4d effective theory, 
the extra factor $\onehalf$  would appear, which 
may be understood as a result of the mixing among $\nu_{R}'$ and $\eta_L^c$.

In this paper the $SO(11)$ gauge-Higgs grand unification has been formulated in the six-dimensional 
hybrid warped space.  The seesaw mechanism for neutrinos naturally emerges,
whose structure is characterized by a $3 \times 3$ mass matrix.
Details, concerning the spectrum of quarks and leptons, evaluation of $V_\eff (\theta_H)$, 
and dynamical EW symmetry breaking, 
will be reported separately.

\begin{acknowledgments}
We would like to thank T.\ Yamashita for valuable comments.
This work was supported in part by Japan Society for the Promotion of Science, Grants-in-Aid 
for Scientific Research,  No.\ 15K05052.
\end{acknowledgments}

\def\jnl#1#2#3#4{{#1}{\bf #2},  #3 (#4)}

\def\Zphys{{\em Z.\ Phys.} }
\def\jssc{{\em J.\ Solid State Chem.\ }}
\def\jpsJ{{\em J.\ Phys.\ Soc.\ Japan }}
\def\ptps{{\em Prog.\ Theoret.\ Phys.\ Suppl.\ }}
\def\PTP{{\em Prog.\ Theoret.\ Phys.\  }}
\def\PTEP{{\em Prog.\ Theoret.\ Exp.\  Phys.\  }}
\def\JMP{{\em J. Math.\ Phys.} }
\def\NPB{{\em Nucl.\ Phys.} B}
\def\NP{{\em Nucl.\ Phys.} }
\def\PLB{{\it Phys.\ Lett.} B}
\def\PL{{\em Phys.\ Lett.} }
\def\PRL{\em Phys.\ Rev.\ Lett. }
\def\PRB{{\em Phys.\ Rev.} B}
\def\PRD{{\em Phys.\ Rev.} D}
\def\PRe{{\em Phys.\ Rep.} }
\def\AP{{\em Ann.\ Phys.\ (N.Y.)} }
\def\RMP{{\em Rev.\ Mod.\ Phys.} }
\def\ZPC{{\em Z.\ Phys.} C}
\def\SCI{\em Science}
\def\CMP{\em Comm.\ Math.\ Phys. }
\def\MPLA{{\em Mod.\ Phys.\ Lett.} A}
\def\IJMPA{{\em Int.\ J.\ Mod.\ Phys.} A}
\def\IJMPB{{\em Int.\ J.\ Mod.\ Phys.} B}
\def\EPJC{{\em Eur.\ Phys.\ J.} C}
\def\PR{{\em Phys.\ Rev.} }
\def\JHEP{{\em JHEP} }
\def\JCAP{{\em JCAP} }
\def\cmp{{\em Com.\ Math.\ Phys.}}
\def\JPA{{\em J.\  Phys.} A}
\def\JPG{{\em J.\  Phys.} G}
\def\NJP{{\em New.\ J.\  Phys.} }
\def\CQG{\em Class.\ Quant.\ Grav. }
\def\ATMP{{\em Adv.\ Theoret.\ Math.\ Phys.} }
\def\ibid{{\em ibid.} }

\def\reftitle#1{}                



\begin{thebibliography}{99}


\bibitem{YH1}
Y.~Hosotani,
\reftitle{Dynamical mass generation by compact extra dimensions}
\jnl{\PLB}{126}{309}{1983};
\reftitle{Dynamics of nonintegrable phases and gauge symmetry breaking}
\jnl{\AP}{190}{233}{1989}.

\bibitem{Davies1}
  A.~T.~Davies and A.~McLachlan,
\reftitle{Gauge group breaking by Wilson loops}
\jnl{\PLB}{200}{305}{1988};
\reftitle{Congruency class effects in the Hosotani model}
\jnl{\NPB}{317}{237}{1989}.


\bibitem{Hatanaka1998}
H.\ Hatanaka, T.\ Inami and C.S.\ Lim,
\reftitle{The gauge hierarchy problem and higher dimensional gauge theories}
\jnl{\MPLA}{13}{2601}{1998}.


\bibitem{ACP2005}
K.~Agashe, R.~Contino and A.~Pomarol,
\reftitle{The Minimal Composite Higgs Model}
\jnl{\NPB}{719}{165}{2005}.

\bibitem{Medina2007}
A.~D. Medina, N.~R. Shah, and C.~E. Wagner,
\reftitle{Gauge-Higgs Unification and Radiative Electroweak Symmetry Breaking 
in Warped Extra Dimensions}
\jnl{\PRD}{76}{095010 }{2007}.


\bibitem{HS2007}
Y.~Hosotani and Y.~Sakamura,
\reftitle{Anomalous Higgs couplings in the $SO(5) \times U(1)_{B-L}$ gauge-Higgs 
unification in warped spacetime}
\jnl{\PTP}{118}{935}{2007};
Y.~Hosotani, K.~Oda, T.~Ohnuma and Y.~Sakamura,
\reftitle{Dynamical Electroweak Symmetry Breaking in $SO(5) \times U(1)$
Gauge-Higgs Unification with Top and Bottom Quarks}
\jnl{\PRD}{78}{096002}{2008}; 
{\it Erratum}-\jnl{\ibid}{{\rm D}79}{079902}{2009};
Y.\ Hosotani, S.\ Noda and N.\ Uekusa,
\reftitle{The Electroweak gauge couplings in $SO(5) \times U(1)$ gauge-Higgs unification}
\jnl{\PTP}{123}{757}{2010}.


\bibitem{FHHOS2013}
S.~Funatsu, H.~Hatanaka, Y.~Hosotani, Y.~Orikasa,  and T.~Shimotani,
\reftitle{Novel universality and Higgs decay $H \go \gamma \gamma, gg$ 
in the $SO(5) \times U(1)$ gauge-Higgs unification}
\jnl{\PLB}{722}{94}{2013};
\reftitle{LHC signals of the $SO(5)\times U(1)$ gauge-Higgs unification}
\jnl{\PRD}{89}{095019}{2014};
\ignore{
\reftitle{Dark matter in the $SO(5)\times U(1)$ gauge-Higgs unification}
\jnl{\PTEP}{2014}{113B01}{2014}.
}
%
S.\ Funatsu, H.\ Hatanaka and Y.\ Hosotani,
\reftitle{$H \go Z \gamma$ in the gauge-Higgs unification}
\jnl{\PRD}{92}{115003}{2015};
%
S.\ Funatsu, H.\ Hatanaka,  Y.\ Hosotani  and Y.\ Orikasa,
\reftitle{Collider signals of $W'$ and $Z'$ bosons  in the gauge-Higgs unification}
\jnl{\PRD}{95}{035032}{2017};
\reftitle{Distinct signals of the gauge-Higgs unification in $e^+e^-$ collider experiments}
arXiv:1705.05282 [hep-ph].


\bibitem{Burdman2003}
G.\ Burdman and Y.\ Nomura,
\reftitle{Unification of Higgs and gauge fields in five dimensions}
\jnl{\NPB}{656}{3}{2003};
%
N.\ Haba, Masatomi Harada, Y.\ Hosotani and Y.\ Kawamura,
\reftitle{Dynamical rearrangement of gauge symmetry on the orbifold $S^1/Z_2$}
\jnl{\NPB}{657}{169}{2003}; 
{\it Erratum}-\jnl{\ibid}{{\rm B}669}{381}{2003};
N.\ Haba, Y.\ Hosotani, Y.\ Kawamura and T.\ Yamashita,
\reftitle{Dynamical symmetry breaking in gauge Higgs unification on orbifold}
\jnl{\PRD}{70}{015010}{2004}.

\bibitem{Lim2007}
C.S. Lim and N.\ Maru,
\reftitle{Towards a realistic grand gauge-Higgs unification}
\jnl{\PLB}{653}{320}{2007}.

\bibitem{Yamashita2011}
K.\ Kojima, K.\ Takenaga and T.\ Yamashita, 
\reftitle{Grand gauge-Higgs unification}
\jnl{\PRD}{84}{051701(R)}{2011};
\reftitle{Gauge symmetry breaking patterns in an SU(5) grand gauge-Higgs unification}
\jnl{\PRD}{95}{015021}{2017}.


\bibitem{Serra2011}
M.\ Frigerio, J.\ Serra and A.\ Varagnolo,
\reftitle{Composite GUTs: models and expectations at the LHC}
\jnl{\JHEP}{1106}{029}{2011};
%
K.\ Yamamoto, 
\reftitle{The formulation of gauge-Higgs unification with dynamical boundary conditions}
\jnl{\NPB}{883}{45}{2014};
%
F.~J.~de Anda,
\reftitle{Left-right model from gauge-Higgs unification with dark matter}
\jnl{\MPLA}{30}{1550063}{2015}.

\bibitem{HY2015a}
Y.\ Hosotani and N.\ Yamatsu, 
\reftitle{Gauge-Higgs grand unification} 
\jnl{\PTEP}{2015}{111B01}{2015}.


\bibitem{Yamatsu2016a}
N.\ Yamatsu, 
\reftitle{Gauge coupling unification in gauge-Higgs grand unification}
\jnl{\PTEP}{2016}{043B02}{2016};
\reftitle{Special Grand Unification}
arXiv: 1704.08827 [hep-ph].  


\bibitem{Furui2016} 
A.~Furui, Y.~Hosotani and N.~Yamatsu,
\reftitle{Toward realistic gauge-Higgs grand unification}
\jnl{\PTEP}{2016}{093B01}{2016}.



\bibitem{Peskin1998}
E.A.\ Mirabelli and M.E.\ Peskin,
\reftitle{Transmission of supersymmetry breaking from a four-dimensional boundary}
\jnl{\PRD}{58}{065002}{1998}.

\bibitem{Minkowski1977}
P.\ Minkowski,
\reftitle{$\mu \go e \gamma$ at a rate of one out of $10^9$ muon decays?}
\jnl{\PLB}{67}{421}{1977}.


\end{thebibliography}
\end{document}